# Time$^2$: A framework for the neural dynamics of visual perception

Laurent Caplette[1,2,3*] & Frédéric Gosselin[2]

[1]Department of psychology, Yale University, New Haven, CT, U.S.A.

[2]Department of psychology, Université de Montréal, Montréal, QC, Canada

[3]CHU Sainte-Justine Azrieli Research Center, Montréal, QC, Canada

*Corresponding author. Email: laurent.caplette@umontreal.ca

**Abstract**

Whenever we look at an object, we seem to perceive it immediately. However, this is not the case for two reasons. First, it takes hundreds of milliseconds for the brain to process visual information reaching the retina. Second, we have to look at an object for a certain amount of time to perceive it (and we typically look at it for hundreds of milliseconds) – during that time, visual information is continuously received on our retinas. These facts together imply that visual information is both processed and received through time. These two temporal facets of perception, which we term processing time and stimulus time, are often conflated in the literature. Moreover, processing time and stimulus time are usually not considered together in experiments. Here, we argue that, to obtain a more complete portrait of visual perception and constrain further models of vision, it is essential to consider and measure both temporal facets simultaneously. We present a new method designed to do so that is based on reverse correlation: *Time²*. We show that this method allows us to precisely characterize many neural phenomena, including rhythmic perception, predictive processing and coarse-to-fine sampling.

**Processing time and stimulus time**

Whenever we look at an object, we seem to instantaneously perceive it and recognize it. However, this is not the case. First of all, our brain takes some time to process the visual input. Light reflected on this object hits our retinas, is transduced into electrical currents, and this electrical activity is transformed in various ways along the cortical hierarchy until a percept is formed and recognition is achieved. This information processing takes some time (around 100-150 ms in most cases; Bullier, 2001; Thorpe et al., 1996) (Figure 1a).

Although this is less discussed, visual information from the object is also continuously being received on our retinas during that time (see Caplette et al., 2020; Caplette et al., 2023). Indeed, because of several factors (such as neural noise and a limited processing capacity), we need to look at stimuli (even static ones) for more than an instant to recognize them, at least for a few tens of milliseconds (Bacon-Macé et al., 2005; Keysers & Perrett, 2002), and sometimes for several hundreds of milliseconds (Henderson, 2003; Hsiao & Cottrell, 2008).

Visual information is thus both *received* and *processed* continuously through time. We refer to these two temporal facets of perception as *stimulus time* and *processing time*. Most studies have considered only one of these facets. Several focused on processing time and studied it using either response times (RT) (Dehaene, 1993; Fabre-Thorpe et al., 1998) or electrophysiological markers (Bullier, 2001; Cichy et al., 2014; Mazer et al., 2002; Siegel, Buschman & Miller, 2015; VanRullen & MacDonald, 2012). Although some of those studies used dynamic stimuli, they did not consider the stimulus time dimension explicitly. For example, VanRullen & MacDonald (2012) presented a disc randomly fluctuating in luminance on every frame, recorded concomitant EEG activity and computed the cross-correlation between both signals to obtain the average response to increases in luminance, which oscillates around 10 Hz (Figure 1b). Despite the stimuli being dynamic, the obtained time course represents processing: the stimulus time dimension is collapsed across when computing the average response. Mazer et al. (2002)

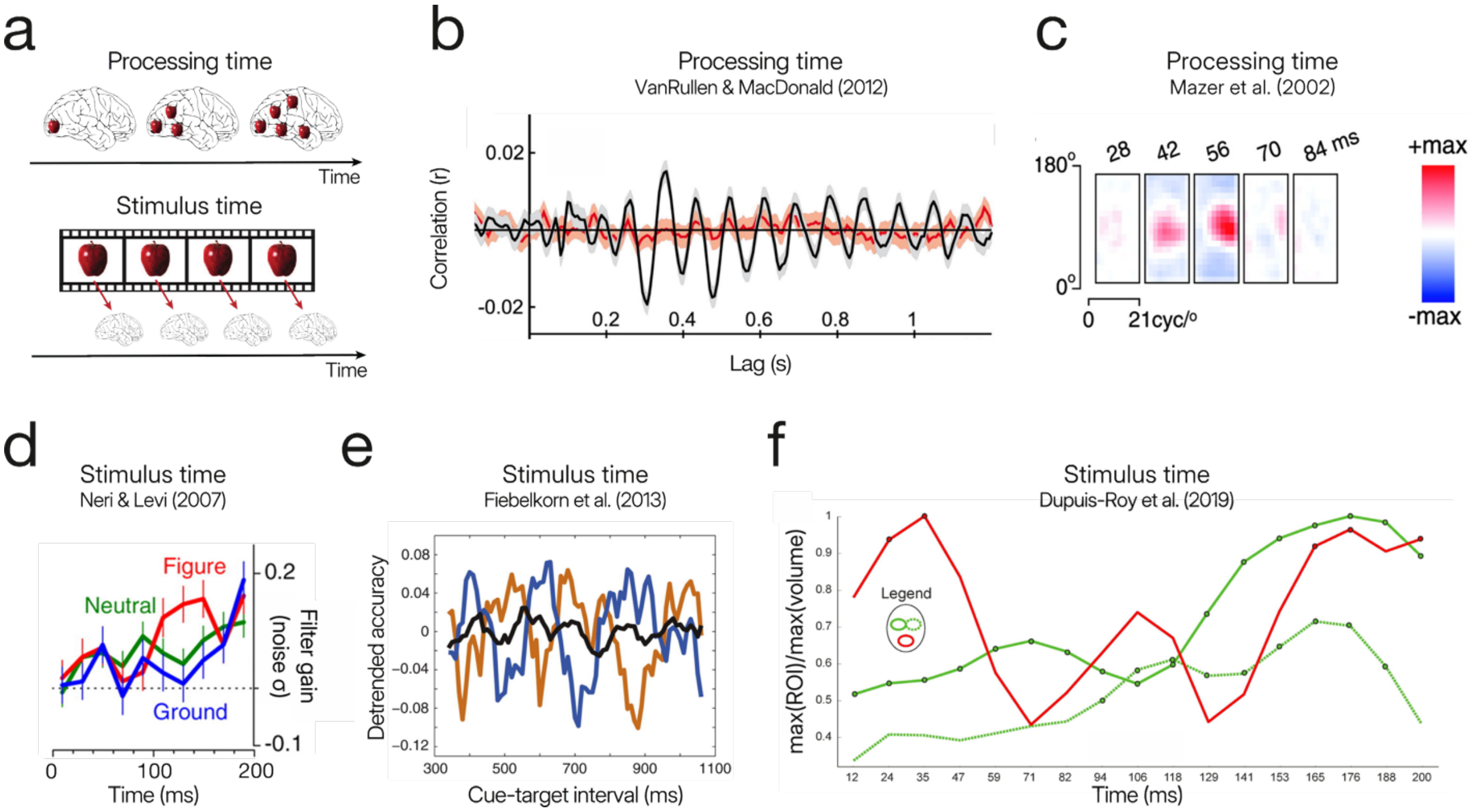

**Figure 1.** Processing time vs stimulus time, and examples of studies investigating each. **a)** Once visual information is received on the retina, it is processed throughout the brain across time: this is processing time. Stimuli are also presented and fixated for some duration and visual information is continuously received on the retina during that time: this is stimulus time. **b)** In this study, a random luminance sequence was shown to participants while their EEG activity was recorded. The black line represents the cross-correlation between the luminance time series and the EEG time series for a specific channel (the red line represents a surrogate time course obtained by cross-correlating the EEG with luminance sequences from different trials). Each time point refers to a different processing latency, not a specific stimulus time point (processing time). Adapted from VanRullen & MacDonald (2012). Copyright (2012), with permission from Elsevier. **c)** Here, the authors showed dynamic sequences of gratings to two adult monkeys and recorded spikes in V1 neurons. They then reverse correlated spike latencies to the orientations and spatial frequencies of the gratings shown. Color represents the normalized magnitude of neural activity for different latencies, spatial frequencies and orientations. Each time point refers to a specific neural latency, not any stimulus time point (processing time). Adapted from Mazer et al. (2002). Copyright (2002) National Academy of Sciences, U.S.A. **d)** In this study, the authors presented dynamic stimuli made of different parts (figure, ground, and neutral) that were superimposed with random luminance noise varying on every video frame across 200 ms. Reverse correlation analyses were then performed to uncover how luminance increases on every stimulus part and video frame correlate with participants' accuracy at a detection task (filter gain). Different time points refer to specific stimulus time points (stimulus time). Adapted from Neri & Levi (2007). Copyright (2007), with permission from the American Physiological Society. **e)** Here, the authors showed cue stimuli followed by target stimuli presented at various locations. The interval between these stimuli was varied randomly between 300 and 1100 ms in 10-ms increments. The participant's accuracy for each interval and target location (one curve per location) was then assessed and quadratic trends were removed from the obtained curves. Different time points represent specific intervals between stimuli (stimulus time). Adapted from Fiebelkorn et al. (2013). Copyright (2013), with permission from Elsevier. **f)** In this study, the authors revealed random parts of face stimuli at random moments across 200 ms on each trial and participants had to identify the sex of the faces. The visibility of each face feature at each moment was then correlated to accuracy at the task. Normalized curves for each feature are plotted, with small black circles denoting significant time points. Each time point refers to a specific stimulus time point (stimulus time). Adapted from Dupuis-Roy et al. (2019).

showed dynamic sequences of gratings to two adult monkeys and recorded spikes in V1 neurons. They then reverse correlated (more on reverse correlation later) spike latencies to the orientations and spatial frequencies of the gratings, revealing that spikes with increasing latencies were associated to increasingly high spatial frequencies but similar orientations (Figure 1c). Because a given time point refers to the average latency of the neural response with respect to the presentation of a grating, the temporal dimension being considered here is also processing time.

Other studies focused on stimulus time by varying the duration of stimuli (Grill-Spector & Kanwisher, 2005; Teichner & Krebs, 1972), presenting different objects or features in different orders (Kauffmann, Chauvin, Pichat & Peyrin, 2015; Parker et al., 1992), manipulating stimulus onset asynchrony in fine increments (Fiebelkorn et al., 2013; Landau & Fries, 2012) or randomly sampling information across time (Blais et al., 2012, 2013; Caplette et al., 2016; Dupuis-Roy et al., 2019; Neri & Heeger, 2002; Neri & Levi, 2007; Vinette, Gosselin & Schyns, 2004), and recording behavioral responses. Neri & Levi (2007) showed stimuli composed of a figure on a background, with random luminance superimposed and varying dynamically across time; they then reverse correlated luminance values at every moment and stimulus part with participants' accuracy at a detection task. They observed that luminance values around 160 ms after stimulus onset correlated most with accuracy (Figure 1d). The temporal facet being considered here is stimulus time, since different time points refer to different instants of the stimulus. Similarly, Fiebelkorn et al. (2013) showed cue stimuli followed by target stimuli presented at various locations, with cue-target intervals varying randomly between 300 and 1100 ms in 10-ms increments, and the participant's accuracy for each interval and target location was assessed. Oscillations in the theta range can be observed in the resulting time series (Figure 1e). Again, time refers here to stimulus time because no neural response is analyzed and time points are purely intervals between two visual events. Finally, Dupuis-Roy et al. (2019) revealed random parts of face stimuli at random moments across 200 ms on each trial and participants had to identify the sex of the faces. The visibility of each face feature at each moment was then correlated to the participant's accuracy. Because time refers to distinct moments of the stimulus, the temporal dimension being analyzed here is also stimulus time.

Few studies investigated both temporal dimensions at the same time, which entails analyzing how information received at each successive moment is processed across time in the brain (Caplette et al., 2020; Caplette et al., 2023; see also King & Wyart, 2021). Moreover, these two temporal aspects of perception are often confused in the literature. In several cases, conclusions about processing time have been drawn based on stimulus timing and conclusions related to stimulus time have been drawn based on temporal differences in information processing (see VanRullen, 2011).

Note that the concepts of stimulus time and processing time are related to those of event time and brain time, sometimes discussed in the temporal perception literature (Johnston & Nishida, 2001). However, whereas our terms refer to dimensions (which are in principle independent and can be simultaneously illustrated, as we will see), these existing terms are usually discussed in relation to a specific event, wherein *event time* refers to the moment at which an event happens in the external world and *brain time* refers to the moment at which neural activity represents the event (that is, specific coordinates on a single dimension of time). Moreover, these concepts are different from *subjective* (or *represented*) time, i.e. the perceived time of an event (Hogendoorn, 2022). While it is often assumed that subjective time and brain time are equivalent, both can be experimentally dissociated: for example, different features of an object are typically processed at different speeds, yet are subjectively integrated into a single coherent percept (Arnold, Clifford & Wenderoth, 2001; Hogendoorn et al., 2010; Johnston et al., 2006; Maunsell et al., 1999; Moutoussis & Zeki, 1997). Therefore, subjective time is probably symbolically encoded by a specialized system in the brain (Heron et al., 2012; Hogendoorn, 2022). We will not discuss subjective time further and will focus on objective time, both external (stimulus time) and neural (processing time).

In this piece, we argue that, to obtain a more complete portrait of visual perception and constrain further models of vision, it is essential to consider both temporal facets simultaneously. Doing so allows us to disentangle competing explanations of many neural phenomena and resolve among them. Hence, we propose a general experimental paradigm to visualize how information received at each moment during stimulus presentation is processed across time in the brain. This approach helps unify many disparate lines of research under a common framework: *Time*$^2$. Below, we describe this paradigm and elaborate on how it enables us to characterize various neural phenomena and consider alternative explanations for existing results.

## Disentangling stimulus time and processing time in the brain

As shown by Wiener, noise can be used to analyze the behavior of any system. This principle underlies reverse correlation studies, in which some stimulus space is randomly sampled and behavioral (Ahumada, 1996; Caplette & Turk-Browne, 2024; Dupuis-Roy et al., 2019; Gosselin & Schyns, 2001, 2003; Murray, 2011; Neri & Levi, 2007) or neural (Ince et al., 2016; Jones & Palmer, 1987; Ringach & Shapley, 2004; Smith, Gosselin & Schyns, 2012) responses are recorded. Analyses — typically linear regressions or weighted sums — are then used to estimate how each part of the space is represented or used by subjects to perform a task (Gosselin & Schyns, 2002; Murray, 2011). For example, images can be shown with trial-by-trial noise (either additive or multiplicative), and the relationship between noise intensity and responses can be evaluated for each pixel. This approach has been extended to the temporal domain: images are shown with spatio-temporal noise to estimate how sensory features from each moment are represented or used by participants to resolve a task. Such methods have been applied using a variety of stimuli, feature spaces and tasks (e.g., Blais et al., 2012, 2013; Caplette et al., 2016, 2021; Dupuis-Roy et al., 2019; Neri & Heeger, 2002; Neri & Levi, 2007; Vinette, Gosselin & Schyns, 2004).

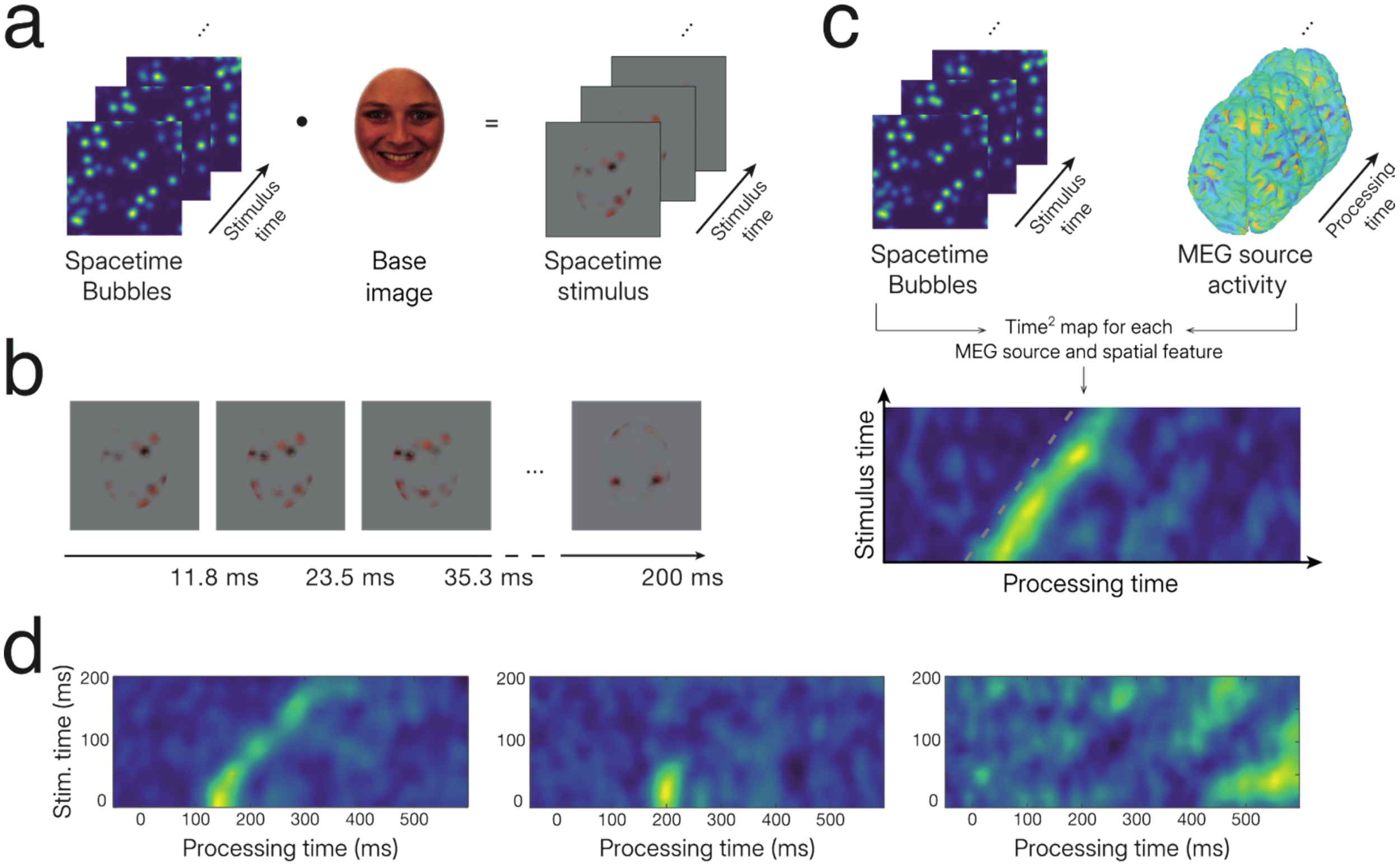

**Figure 2.** Experimental paradigm and analyses. **a)** Creation of stimuli. A three-dimensional array, of which two dimensions are pixel space and one dimension is stimulus time, is filled with three-dimensional Gaussian apertures (Spacetime Bubbles). This noise array is multiplied element-wise with a base image (here, a cropped face), replicated across stimulus time. This results in a spacetime stimulus in which spatial parts of an image are randomly sampled across time. **b)** An example of a spacetime stimulus, shown frame by frame. **c)** Overview of analyses. Brain activity for each MEG source and processing latency is regressed onto Spacetime Bubbles values for each pixel (or the average values of pixels within a spatial feature, such as the mouth area in a face) and stimulus moment (i.e., video frame) across trials. Resulting regression coefficients can be arranged such that one Time$^2$ map is obtained for each MEG source and pixel (or spatial feature). On such a map, stimulus time is illustrated on the y axis, from bottom to top, and processing time is illustrated on the x axis, from left to right. The superimposed dashed line here indicates a slope of 1, i.e. identical latencies irrespective of stimulus moment. **d)** A clustering analysis revealed three types of Time$^2$ maps in Caplette et al. (2023). One representative map of each type is shown here. While the leftmost and middle map types were observed primarily in early visual areas, the rightmost type was found mainly near the lateral occipital complex.

We further extended this approach by jointly considering the temporal aspects of both stimuli—as just explained—and responses. Specifically, we revealed randomly located image features at random moments (by applying multiplicative spatio-temporal noise to the stimuli), and we recorded concurrent temporally resolved brain activity (Figure 2a–b). As described above, we then performed regressions to recover how *each image feature* received on the retina *at each moment* is processed across time at each recorded brain location. This can be represented as a *Time$^2$ map* for each image feature and brain location (Caplette et al., 2020, 2023). In such maps, each row shows the processing, over time, of the image feature received at a specific instant on the retina. Processing can theoretically be observed anywhere in a map except in the region above the slope-of-1 diagonal starting at the origin (this would otherwise violate the arrow of time). For instance, if information reaching the retina at any given time is processed with a constant delay in a brain area, information presented some time later (higher on the *y* axis) will be processed the same amount of time later (further on the *x* axis), resulting in a diagonal activation pattern (Figure 2c). This is not necessarily the case for all brain areas however, especially those higher in the cortical hierarchy (e.g., the lateral occipital complex; Figure 2d). By illustrating stimulus time and processing time simultaneously, Time$^2$ maps allow us to characterize neural information processing while accounting for the fact that stimuli are temporally extended and that visual information is continuously impinging on our retinas.

This method is related to previous work in which researchers showed stimuli with luminance varying randomly over time to participants while recording their EEG activity (Lalor et al., 2006, 2009; VanRullen & MacDonald, 2012; see also Crosse et al., 2016, 2021; Smith & Kutas, 2015). However, these studies used stimuli lasting several seconds and performed cross-correlation analyses to obtain a single neural timecourse representing the average response to the stimulus or feature, averaged across all frames of the stimulus. Here, we obtain a distinct response for each frame. Our method can also be seen as a generalization of time-varying (and lag-varying) connectivity: instead of analyzing correlations between the activity from two ROIs at different time points, we analyze the correlations between the amplitude of stimulus features and the activity in one ROI at different time points. Finally, our paradigm improves upon experimental designs that compare only a limited set of stimulus time courses (e.g., low to high vs high to low spatial frequencies; Fintzi & Mahon, 2014; Kauffmann et al., 2015). Importantly, note that any feature space can be sampled, just as in classical reverse correlation: spatial locations, but also spatial frequencies, orientations, deep neural network features, etc.

We recently applied this method to investigate information sampling and processing during face recognition tasks (facial expression recognition and sex recognition). Our results showed that information received at different moments within an eye fixation is processed differently, with information processing being often far from a simple diagonal. Distinct patterns can be seen across the visual cortex, likely due to a combination of bottom-up and top-down factors, including neural oscillations (Figure 2d; Caplette et al., 2020; Caplette et al., 2023). We elaborate on these findings in the following section to illustrate how this framework can advance our understanding of the neural dynamics of visual perception.

## Time$^2$ allows for a characterization of various neural phenomena

One important reason our method is useful is because stimulus time matters to the brain: information received at different moments on the retina, even for a static stimulus, is processed differently by the brain. This is exemplified by many neural phenomena. For example, low-level mechanisms such as adaptation and repetition priming may reduce or enhance, respectively, the processing of information received later. In addition, information received at different phases of a neural oscillation cycle may be processed to different degrees (Blais et al., 2013; Helfrich et al., 2018; Landau & Fries, 2012; VanRullen, 2016). Different features may also be attended to in a specific order during stimulus presentation to increase computational efficiency (e.g., low vs high SFs; Caplette et al., 2016; Parker et al., 1992; Hegdé, 2008; Watt, 1987). Information received early may be held in a buffer for some time to be integrated with information received later (Hanks & Summerfield, 2017; Miller, 1988), and information received later may be subject to more top-down influences (Bar, 2003; Oliva & Torralba, 2006). Because they depict both processing time and stimulus time simultaneously, Time$^2$ maps are helpful to visualize these phenomena. Model maps can be generated directly from hypotheses about the processing of dynamic inputs or from simulations modelling specific observer characteristics. These model maps can then be compared visually to observed maps obtained from experimental data. A more formal comparison of the results to a model can also be performed (e.g., Caplette et al., 2023). Below, we elaborate on some neural phenomena such as those discussed above (with different hypotheses illustrated on Figure 3).

*Neural oscillations*

Oscillations are highly prevalent in the brain and occur independently of sensory stimulation (Buszaki & Draguhn, 2004). Ongoing endogenous oscillations may influence how information received at different moments is processed: information reaching a brain region at a given moment may be more or less effectively processed depending on the instantaneous phase of local oscillations at that time (Figure 3a, left). Several behavioral studies have reported that perceptual sensitivity and attention oscillate over the duration of a stimulus (Blais et al., 2013; Fiebelkorn et al., 2013; Landau & Fries, 2012; VanRullen, 2016; VanRullen et al., 2007; but see Keitel et al., 2022; van der Werf et al., 2022). For example, the threshold for detecting two successive flashes oscillates at about 33Hz (Latour, 1967), and attention seems to oscillate between monitored stimuli at a frequency of 7-8 Hz (Fiebelkorn et al., 2013, 2018; Holcombe & Chen, 2013; Landau & Fries, 2012). Such *oscillatory* *sampling* has been linked to ongoing brain activity: detection performance has been shown to be modulated by the phase of theta and alpha oscillations shortly before stimulus onset (Busch et al., 2009; Hanslmayer et al., 2013; Helfrich et al., 2018; Fiebelkorn et al., 2018).

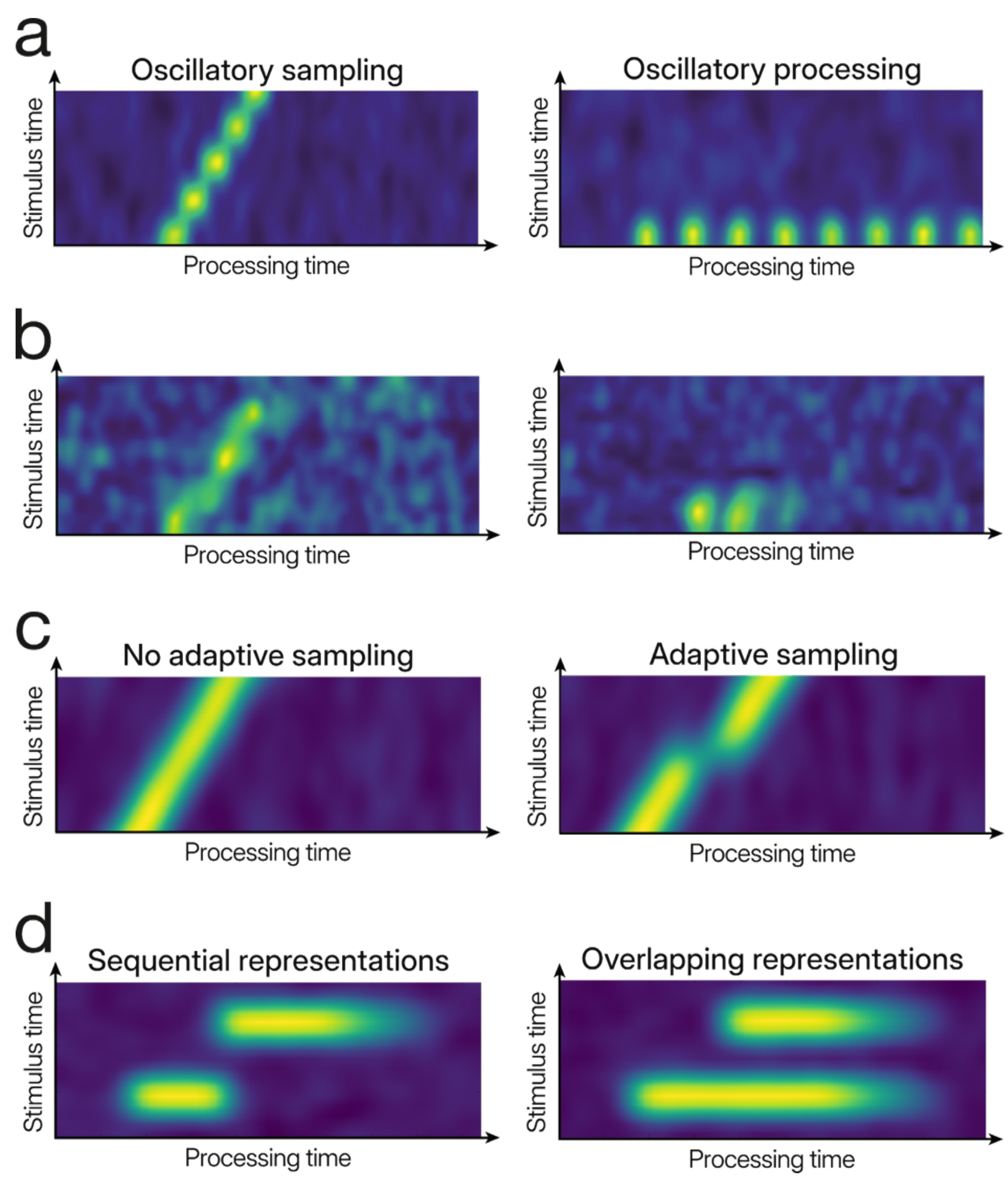

**Figure 3.** Time$^2$ maps enable the visualization and precise characterization of many neural phenomena. **a)** Neural oscillations. Ongoing endogenous oscillations might theoretically interact with continuously incoming visual information in various ways. (Left) This interaction might result in oscillatory sampling, such that information received on the retina at successive moments is more or less processed, in a rhythmic way. (Right) Alternatively, it might result in oscillatory processing, such that information received at a given moment is processed in a rhythmic way across time. **b)** Although we observed both oscillatory patterns in the brain (shown here for two sources in occipito-temporal areas; see Caplette et al., 2023), sampling oscillations were by far the most prevalent. **c)** Predictions. If the base image being shown is changed in the first half of the stimulus on some trials, and processing is adaptive, processing in the latter half of the stimulus could be altered in some way. Here, Time$^2$ maps are shown for these altered trials, for the hypothetical case of no adaptive sampling (processing in the second half is similar; left panel) and adaptive sampling (processing in the second half is different; right panel). **d)** Information maintenance. If two scenes are shown shortly one after another, the representation of the first one could be maintained merely until the representation of the second one comes in, or later, so that representations are overlapping.

Ongoing neural oscillations may also modulate information processing across time (Figure 3a, right). In this case, even a single snapshot of visual information would be processed rhythmically, with the amplitude of related brain activity increasing and decreasing in a periodic manner (*oscillatory processing*). For example, distinct visual features have been shown to be processed at different frequencies (Schyns et al., 2011; Smith et al., 2006), and alpha oscillations seem to implement a reverberation of sensory input that could be responsible for the maintenance of representations over time (VanRullen & MacDonald, 2012).

Until recently, it remained unclear whether oscillatory processing or oscillatory sampling was the more prevalent consequence of endogenous neural oscillations. Using the paradigm described above, we showed that, although both phenomena could be observed, oscillatory processing was much less common, suggesting that the primary manifestation of endogenous oscillations on sensory processing is a modulation of sampling (Caplette et al., 2023) (Figure 3b). We further showed that sampling oscillations occurred across the occipital cortex and beyond, at most frequencies analyzed (7-20 Hz). These results provide insights into how ongoing neural oscillations interact with continuously incoming visual information, even when it is from a single static stimulus.

*Predictions and postdictions*

Predictions are ubiquitous in the brain. Several influential theories of brain function posit that high-level brain areas continually predict activity in lower areas, and that only prediction errors are sent to higher-level areas (Friston, 2005; Friston & Kiebel, 2009; Rao & Ballard, 1999). Recently, these theories have been extended to apply to dynamic stimuli, wherein predictions are updated over time not only through iterative processing loops in the brain but also based on the visual information that is continuously incoming (see de Vries & Wurm, 2023; Jiang & Rao, 2024; Millidge et al., 2024). According to these models, visual information received early is used to modulate the processing of information received subsequently. Early studies have shown that some features received early can indeed constrain the later processing of other features (e.g., early global features constraining later local features; Oliva & Torralba, 2006; Sanocki, 2001). It is, however, unclear whether this *adaptive modulation* occurs within a fixation and what mechanisms might underlie it (Ballard, 2015; Fiebelkorn & Kastner, 2019). Our paradigm can provide an answer to this question. For example, to test whether the visual system uses global information received early to guide the processing of later fine-grained information, we could use a 2-alternative recognition experiment similar to what we have used until now, but in which the base image being sampled in the first half of the temporal stimulus is altered (e.g., to be of the opposite alternative) on half the trials. If the processing of information shown in the second half of the stimulus (which remains identical) is qualitatively different in those altered trials compared to the unaltered trials, modulation is likely occurring. For example, processing in a given brain region could be reduced or delayed if this region is more tuned to the features of the first base image (Figure 3c shows one possibility, i.e. that of a constant processing delay). Our method would further allow us to characterize precisely how and where in the brain this effect occurs.

An even more intriguing possibility is that, in higher-level regions, information received late is processed earlier and influences the subsequent processing of information sampled earlier. This phenomenon, known as postdiction, has been studied extensively. A classic example is backwards masking, in which a masking stimulus presented shortly after a first target stimulus inhibits the processing of the target. This effect may occur due to the presence of buffers in the brain that temporarily store information (Herzog et al., 2020; Shimojo, 2014). It is unknown however whether postdictions occur in the context of ordinary perceptual decisions such as categorizations. A similar experiment to the one discussed above could be devised to test this idea, but in which later information is altered, and the processing of information received earlier (which remains unaltered) is analyzed. Again, if the processing of information received early differs depending on the content of information shown later, this would indicate the presence of postdictive mechanisms.

*Information maintenance*

Information maintenance is a fundamental mechanism that intervenes in the other phenomena discussed in this opinion piece—for example, prediction errors, which require maintaining the past to compute discrepancies, and postdictions, where the past must be sustained for the future to influence it. More generally, visual input—or at least portions of it—must be maintained for the visual system to process it in various ways (e.g., Herzog et al., 2020; McClelland, 1979; Wolff et al., 2022). However, it is unclear how this maintenance is implemented in various brain regions. For example, one possibility is that information is maintained until the visual input changes, at which point all ongoing processing is disrupted and the representation is overwritten by the updated information. This idea was first proposed as a potential explanation for the results of change blindness experiments, in which observers can't detect how two scenes shown consecutively change (Beck & Levin, 2003; O'Regan & Noë, 2001; Rensink et al., 1997). In a Time$^2$ map, this would be visible as two horizontal lines not overlapping in the processing time dimension, with the processing of the first representation stopping as soon as the processing of the second representation starts (Figure 3d, left). More recent work in that area instead proposed that both representations coexist, suggesting that the source of change blindness lies in the process of comparing the representations rather than them being inexistent (Beck & Levin, 2003; Frey et al., 2024; Hollingworth & Henderson, 2002; Smith, Lamont, & Henderson, 2012). In other words, the processing of the new scene would start, with all processing of the previous scene continuing unaffected. This would also be visible in a Time$^2$ map as horizontal lines for each scene, but in this case the lines would be overlapping in the processing time dimension (Figure 3d, right). Another more drastic view is that, at least in brain regions responsible for change detection, internal memory storage does not exist, and rather the world acts as an "outside memory" (O'Regan, 1992; Rensink, O'Regan, & Clark, 1997; Simons & Levin, 1997). This would correspond as a slope-of-1 diagonal in a Time$^2$ map, where information is immediately processed when it arrives and not maintained at all (e.g., Figure 2c). Importantly, these different implementations of information maintenance (or absence of maintenance) are likely to be visible in different brain regions.

### Time$^2$ provides a framework for evaluating alternative explanations of neural latencies

Because visual information is not only processed across time but also received across time, some temporal neural phenomena can be explained in terms of either processing time or stimulus time. Specifically, when a brain region responds to two distinct features with different latencies, this may result either from differences in the speed at which these features were processed prior to that region (their *processing speed*; Figure 4, top) or from differences in when those features were received on the retina (their *stimulus moment*; Figure 4, middle). Although the retina receives all visual information indiscriminately from the external world, higher-level brain regions may receive specific features from specific stimulus moments filtered by earlier areas. Differences in stimulus moment might arise from attentional mechanisms prioritizing certain time windows (Fiebelkorn et al., 2013; VanRullen, 2016), or from other perceptual processes such as those discussed above. Notably, it could also serve as a compensatory mechanism: features processed more slowly might be sampled from earlier stimulus moments, so that they both arrive at a similar time in a higher-level region (Figure 4, bottom). This distinction between processing speed and stimulus moment also applies, to some extent, to behavioral response times (RTs): differences in RTs to various features may reflect differences in either how fast the features are processed or when they are initially sampled.

Such phenomena have been documented. For example, color is processed more slowly than luminance in the early visual cortex (Maunsell et al., 1999; Nowak et al., 1995), yet no such latency difference is observed in the primate higher-level brain areas during categorization tasks involving color and grayscale faces (Edwards et al., 2003). There is some evidence that this could be the result of processing color from earlier stimulus moments to compensate for its slower processing (Dupuis-Roy et al., 2019).

A related case is the well-known *coarse-to-fine* processing sequence in vision: from low to high spatial frequencies (SFs). Again, distinguishing processing speed from stimulus moment is crucial. Both are often discussed (and sometimes conflated) in the literature. Low SFs are known to be processed faster than high SFs by retinal and LGN cells (Allen & Freeman, 2006; Nowak et al., 1995), and they evoke spikes tens of milliseconds earlier in the early visual cortex (Mazer et al., 2002; Purushothaman et al., 2014) — a phenomenon we call *coarse-to-fine processing*. In parallel, other studies have reported that processed low SFs tend to originate from earlier stimulus moments than high SFs, suggesting that low SFs are attended before high SFs during stimulus fixation (Caplette et al., 2016; Parker et al., 1992; Schyns & Oliva, 1994) — this can be termed *coarse-to-fine sampling*. While both phenomena seem to coexist, coarse-to-fine sampling is likely more flexible and top-down dependent (Caplette et al., 2021; Oliva & Schyns, 1997; Schyns & Oliva, 1999), making it a more plausible mechanism for efficient scene recognition (Marr, 1982; Watt, 1987) and top-down facilitation (Bar, 2003; Bullier, 2001).

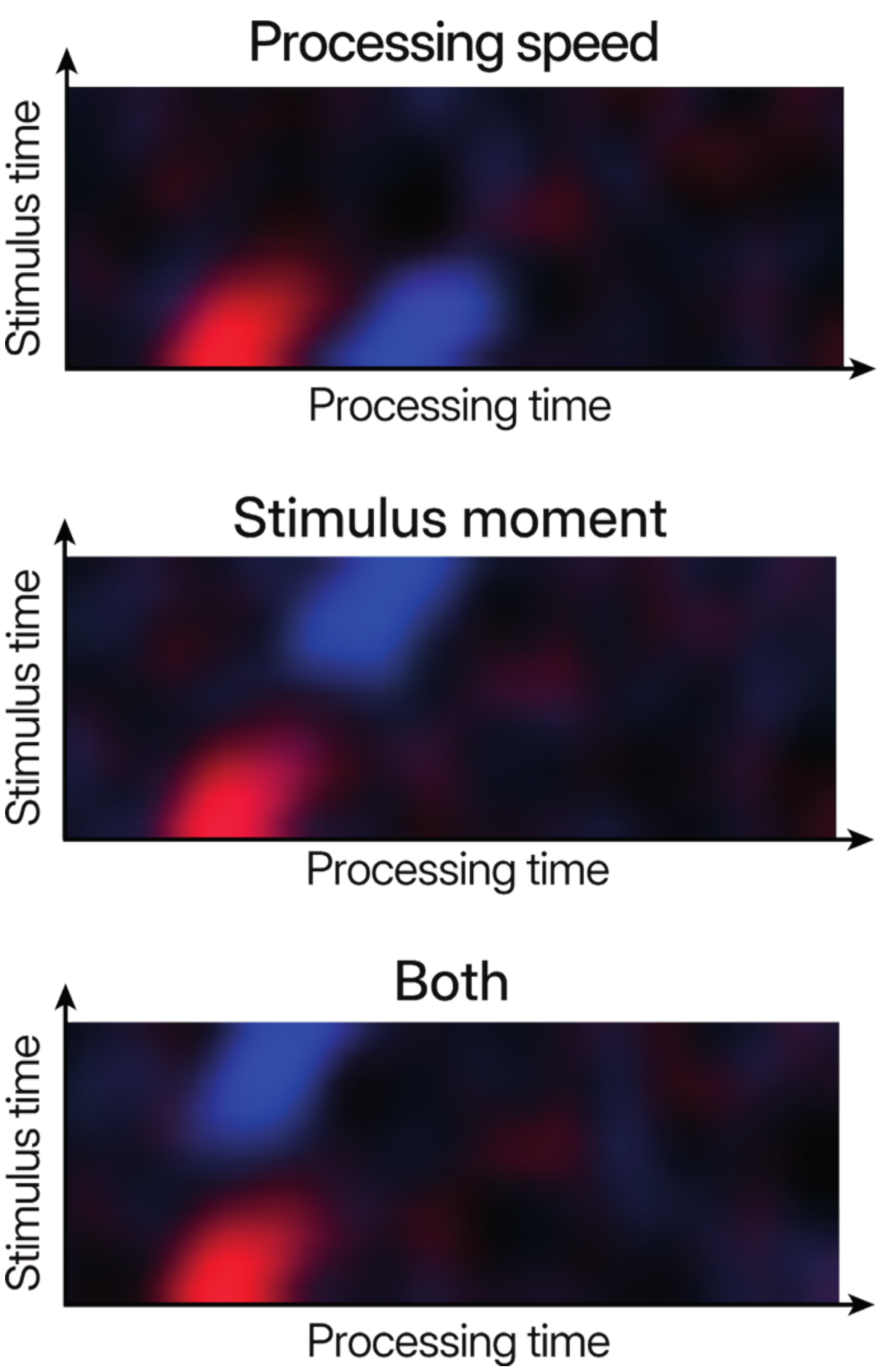

**Figure 4.** Examples of how a feature's processing speed and its stimulus moment can influence when it is processed in a higher-level brain region. The processing of two distinct features is represented by the saturation of the red and blue color channels. (Top) Two different features received at the same moment are processed at different speeds and so their latency in a higher-level brain region is different. (Middle) Two different features are processed at the same speed (i.e., are on the same slope-of-1 diagonal), but from different stimulus moments in a high-level brain region: this also results in different latencies in that region (different positions on the *x* axis). (Bottom) Features are both processed at different speeds and from different stimulus moments, here in such a way that their latency in a high-level brain region is about the same.

Beyond those discussed above, many other features of various sensory modalities are processed at different speeds (e.g., auditory vs visual information, low-contrast vs high-contrast images, color vs motion; Burr & Alais, 2006; Eagleman, 2010; Moutoussis & Zeki, 1997; Wolfe, 2014). These differences pose challenges for *temporal binding*, i.e. the integration of asynchronously processed features into a unified percept (Hogendoorn, 2022; Zivony & Eimer, 2022, 2024). Processing different features from distinct stimulus moments may help the brain solve this problem (Hogendoorn & Burkitt, 2019).

## Concluding remarks

We presented an experimental framework to disentangle stimulus time and processing time in the brain and unify various temporal phenomena under a single conceptual umbrella. We believe this paradigm holds promise for investigating rhythmic perception, predictive processing, information maintenance and other related phenomena. For instance, it enabled us to characterize how ongoing oscillations interact with continuously incoming visual information (Caplette et al., 2023). In the future, this paradigm could be extended in a number of ways. For example, our temporal sampling method could be applied to intrinsically dynamic stimuli (e.g., facial expressions; Blais et al., 2017) and extended to longer time scales along both the stimulus time and processing time dimensions, to more realistically capture temporal dynamics. Furthermore, our paradigm offers an opportunity to test and refine models and theories of temporal perception. For example, recurrent neural networks have increasingly been proposed as promising models of the visual system (Güçlu & van Gerven, 2017; King & Wyart, 2021; van Bergen & Kriegeskorte, 2020). These models implicitly assume that information reception and processing occur simultaneously over time, but whether they sample information across time in the same way as humans remains to be tested. In addition to testing computational models of perception, our paradigm enables testing various theories of temporal perception, across fields such as attention, visual search, temporal integration, evidence accumulation, and others. Beyond our paradigm, other decoding-based approaches have been useful to get insights into the processing dynamics of different events happening successively in time (King et al., 2021; Subramaniyan et al., 2018; Turner et al., 2025). Similar multivariate decoding could also be integrated into our approach to target representational content instead of pure neural activation. Moreover, obtaining perceptual judgments would allow us to relate processing and stimulus time to a third type of time: the perceived timing of different events, i.e. *subjective time* (Hogendoorn, 2022; Johnston & Nishida, 2001). More generally, we hope that our work will help the broader neuroscience community recognize the important distinctions between different types of time, and we encourage researchers to be more explicit about what they are measuring and more careful in interpreting their results.